\documentclass[amsfonts,10pt]{article}
\usepackage{graphicx}
\usepackage{amsmath}
\let \bbox=\bf
\begin{document}
\centerline{\LARGE  Master-equations for the study of
decoherence}\par
\vskip 15  pt
\centerline{B.~Vacchini\footnote{Dipartimento di Fisica
dell'Universit\`a di Milano and Istituto Nazionale di Fisica
Nucleare, Sezione di Milano, Via Celoria 16, I-20133, Milan,
Italy. E-mail: bassano.vacchini@mi.infn.it}}
\begin{abstract}
   Different structures of master-equation used for the description of
   decoherence of a microsystem interacting through collisions with a
   surrounding environment are considered and compared. These results
   are connected to the general expression of the generator of a
   quantum dynamical semigroup in presence of translation invariance
   recently found by Holevo.
\end{abstract}
\section{Introduction}
\label{sec:introduction}
In recent times the word decoherence has become quite fashionable in
order to describe a range of utterly different physical situations,
which however all exhibit a common qualitative feature: a quantum
system, due to its unavoidably imperfect isolation from the
surrounding environment, shows in its dynamical evolution the
suppression of typical quantum coherence properties, such as
interference capability. Although the subject is in rapid evolution, a
nice recent presentation of the field, anchored to the robust
background of the theoretical description of open quantum systems, can
be found in~\cite{Petruccione}.
\par
The basic ideas are actually very old and as it was recently stressed
in~\cite{Dube} can be essentially traced back to the first studies on
the measurement problem in quantum mechanics in the 50's. These
studies in which the main concepts related to decoherence already
appeared has led by now to relevant improvements in the formulation of
quantum mechanics, going beyond Dirac's presentation and leading to the
new concept of effect, positive operator valued measure, operation and
instrument, also disvealing most fruitful and interesting connections
with the theory of stochastic processes; it appears indeed that many
useful clues in studying the theory of quantum systems can be obtained
from classical probability theory, rather than from the usual
correspondence with classical mechanics. These more advanced and
flexible tools in the description of quantum systems and their
dynamics are by now extensively used in quantum information and
communication theory.
\par
Indeed the common root between decoherence and the measurement problem,
that is the interaction between a microsystem and a macrosystem, obviously
indicates that concepts, techniques and tools originating in the realm
of foundations of quantum mechanics will prove an essential ingredient
in the actual study of decoherence.
In this respect a relevant distinction is to be pointed out. One thing
is the loss of quantum coherence for a microsystem interacting with some
macrosystem, another thing the classical behavior which macrosystems
actually exhibit, thus allowing for an objective description. Though
the latter phenomenon can be thought of as a consequence of the first,
enhanced by the huge number of degrees of freedom pertaining to a
macrosystem, the two physical situations are actually very different
and it might well be the case that different approaches should be
devised, the connection being not necessarily trivial as is often
conjectured or implicitly assumed.  In particular the description of
macrosystems should rely on a suitable development of quantum
statistical mechanics, which extended to non-equilibrium situations
could allow for the appearance of a classical behavior for a subset of
observables, possibly giving useful insights in the description of decoherence
for a microsystem~\cite{torun99-torun01}. Since the connection between the
phenomenon of decoherence and the measurement problem has been touched upon, it
is important to stress that decoherence is not a solution to the aforementioned
problem. This incorrect viewpoint is often implicitly or explicitly
assumed, however as recently most clearly shown in the standard
framework of Dirac's formulation of quantum
mechanics~\cite{Bassi-decoh}, even supposing that due to interaction
with the environment the combined system composed by microsystem and apparatus would end
up in a statistical mixture with respect to some pointer basis, there
is no reason for this basis to be factorisable and typically in the
case of a measurement apparatus the combined system will not exhibit
macroscopically distinct states.
\par
The reason for the recent renewal of interest in decoherence is twofold: first
enormous experimental progress has been made in dealing in a
controlled way with microsystem, also engineering superposition states which
might be particularly sensitive to decoherence effects, and second decoherence is perhaps
the worst enemy when it comes to the physical implementation of
quantum computers. These two major motivations push current research
work in the two related directions of both quantitatively
understanding and avoiding decoherence.  Up to now most theoretical models of decoherence
have been chosen rather because of their solubility, than because of
their adherence to realistic physical models. The very universality
which is often expected and advocated for the phenomenon actually
relies on suitable modeling for reservoir and interaction, so that
specific properties of system, bath and interaction should be of
relevance in explaining the sensitivity to the different environmental
couplings which actually appears in experiments, as recently stressed
for example in~\cite{Dube,HuPRA-AnastopoulosIJTP}. Real progress in
modeling and understanding of the phenomenon depends on a detailed
description of the physical system and of its dynamics. In this spirit
in the following we will try to focus on structures of master-equation
(ME) which apply to the description of decoherence of a neutral massive microsystem
coupled to the environment due to collisions with environmental particles. Many
proposals have been put forward in the literature and we will briefly
outline the relationships among the different models. The Markovian
description level pertaining to these ME is certainly not the most
general physical picture, but seems appropriate to this kind of
dynamics. It is of course of particular interest to investigate how
and under which conditions ME do emerge from a more refined
description of the reduced dynamics of the microsystem, as has recently been
done in~\cite{AnkerholdLNPH} with reference to the path integral
approach, where non-Markovian effects and strong coupling can be taken
into account, but the noise is essentially bound to be Gaussian.
\par
Useful insights in the structure of the Markovian ME can be gained
from the mathematical characterization of generators of quantum
dynamical semigroups~\cite{Fannes-AlickiPL} recently given by Holevo
under the further requirement of translation invariance
(TI)~\cite{HolevoRAN-HolevoJMP}. This characterization arises from a deep
analogy with the classical L\'evy-Khinchin formula for the
characteristic function of a L\'evy process, i.e. a stochastic process
homogeneous in space and time, thus having independent and stationary
increments; it provides a more detailed description for the possible
structure of the generator of the dynamics than the Lindblad result.
\par
The paper is organized as follows: in
Sec.~\ref{sec:models-decoh-induc} we deal with different ME for the
description of decoherence induced by collisions; in Sec.~\ref{sec:covar-prop}
we compare these results with the structures arising in presence of
TI.
\section{Models of decoherence induced by collisions}
\label{sec:models-decoh-induc}
A useful model for the description of decoherence was first obtained
in~\cite{Joos-Zeh}
\begin{equation}
   \label{eq:1}
         {  
        d {\hat \rho}  
        \over  
                      dt
        }  
        =
        -
        {i \over \hbar}
        [{\hat H}
        ,
        {\hat \rho}
        ]
        +
        JZ [\hat \rho]  
\qquad
        JZ [\hat \rho]  
        =
        -\Lambda 
        \sum_{i=1}^3
        \left[  
        {\hat{\bbox{x}}}_i ,
        \left[  
        {\hat{\bbox{x}}}_i ,\hat \rho
        \right]  
        \right],  
\end{equation}
where $\hat \rho$ is the statistical operator associated to the
microsystem and ${\hat H}$ its Hamiltonian which here and in the
sequel we take to be that of a free particle. This ME
describes the dynamics of the center of mass of the microsystem
due to scattering with an incoming particle flux. It  allows in
a straightforward way for the introduction of a typical decoherence
time after which off-diagonal matrix elements of the
statistical operator in the position representation are heavily
suppressed. It is in particular seen as the recoilless limit of the
so-called Caldeira Leggett ME
\begin{equation}
   \label{eq:2}
        CL [\hat \rho]  
        =
        -\gamma 
        \frac{2M}{\beta\hbar^2}\sum_{i=1}^3
        \left[  
        {\hat{\bbox{x}}}_i ,
        \left[  
        {\hat{\bbox{x}}}_i ,\hat \rho
        \right]  
        \right]       -
        {i\over\hbar}
        \gamma
        \sum_{i=1}^3
        \left[  
        {\hat{\bbox{x}}}_i ,
        \left \{  
         {\hat{\bbox{p}}}_i ,\hat \rho
        \right \}  
        \right],            
\end{equation}
where $\gamma$ is the friction coefficient, $\beta$ the inverse
temperature and $M$ the mass of the microsystem. Eq.~(\ref{eq:2}) is
usually considered for the description of quantum Brownian motion in
the high temperature limit, which is not necessarily always the case
in experimental setups where one wants to investigate coherence
properties of the system and their washing out due to controlled or
uncontrolled coupling to the external environment. The high
temperature limit is linked to the fact that~(\ref{eq:2}), in contrast
with~(\ref{eq:1}), cannot be cast into Lindblad form and therefore
does not preserve positivity of the time evolution.  A further term of
the form $- \chi\gamma \frac{\beta}{M} \sum_{i=1}^3 \left[
  {\hat{\bbox{p}}}_i , \left[ {\hat{\bbox{p}}}_i , \hat \rho\right]
\right] $ with $\chi \geq \frac{1}{8}$ is necessary in order to
preserve complete positivity~\cite{SandulescuIJMPE,art5}, a term which
due to its different $\beta$ dependence can be neglected in the high
temperature limit. Different values of the coefficient $\chi$ have
appeared in different models~\cite{DiosiEL93-DiosiP,DiosiEL95}, but it
appears that the correct value should be the \textit{minimal}
correction $\chi = \frac{1}{8}$. The friction term in~(\ref{eq:2})
accounts for energy transfer and therefore thermalization of the
Brownian particle, leading to the
existence of a stationary solution of the form $e^{-\beta { {\hat
      {\bbox{p}}}^2 \over 2M } }$; though the typical time scales for decoherence and
relaxation in this kind of models may easily differ by orders of
magnitude, so that thermalization takes place on a much longer time
scale, it is nevertheless of interest to consider a possibly fully
realistic description, where all physical processes can be correctly
described. In fact as has been pointed
out~\cite{Ballentine-Gallis91} the ME~(\ref{eq:1}) predicts a steady
growth in energy for the microsystem. Shortly after the
proposal~(\ref{eq:1}) another ME
\begin{equation}
   \label{eq:5}
        GRW [\hat \rho]  
        =
        -\lambda
        \left (\hat \rho -
         \sqrt{\frac{\alpha}{\pi}}
        \int 
        d^3 \!
        {\bbox{s}}
        \,  
         e^{-\frac{\alpha}{2} (\hat{\bbox{x}}-\mathbf{s})^2}
         \hat \rho
         e^{-\frac{\alpha}{2} (\hat{\bbox{x}}-\mathbf{s})^2}
        \right ) 
\end{equation}
has been introduced by Ghirardi, Rimini and Weber~\cite{GRW}, which
also predicts a steady grow in energy (though for the proposed values
of the parameters $\alpha$ and $\lambda$ the growth is actually by
orders of magnitude insignificantly small). The result~(\ref{eq:5})
stands however on a completely different footing, since it is not
meant as an appropriate description of the dynamics of a microsystem
interacting with some environment, but as a fundamental modification of
Schr\"odinger's equation allowing to solve the measurement problem and
can be obtained from the latter by the insertion of a stochastic
correction corresponding to white noise (for a recent review
see~\cite{GRW-review}).
\par
The result of Joos and Zeh was later improved by Gallis and
Fleming~\cite{Gallis90}, always neglecting recoil effects. The
motivation for this further work was the observation that due
to~(\ref{eq:1}) the incoherent part of the time evolution induces a
suppression of the off-diagonal matrix elements according to
$\frac{\partial}{\partial t}\langle
\mathbf{x}|\hat\rho|\mathbf{y}\rangle=
-\Lambda|\mathbf{x}-\mathbf{y}|^2\langle
\mathbf{x}|\hat\rho|\mathbf{y}\rangle$, where the localization factor
grows without bound for $|\mathbf{x}-\mathbf{y}|$ going to infinity.
On physical grounds it is expected that such a behavior might hold at
short length-scale, i.e. small $|\mathbf{x}-\mathbf{y}|$, while for
long length-scale there should be no dependence on the spatial
separation, otherwise the environment would have to be self-correlated
over an infinite length scale. This unphysical feature does not appear
in the model of quantum mechanics with spontaneous localization, in
fact according to~(\ref{eq:5}) one would have
\begin{equation}
   \label{eq:7}
   \frac{\partial}{\partial t}\langle
   \mathbf{x}|\hat\rho|\mathbf{y}\rangle=
-\lambda\frac{\alpha}{4}|\mathbf{x}-\mathbf{y}|^2\langle \mathbf{x}|\hat\rho|\mathbf{y}\rangle
\qquad \mathrm{and} \qquad
   \frac{\partial}{\partial t}\langle
   \mathbf{x}|\hat\rho|\mathbf{y}\rangle=
-\lambda\langle \mathbf{x}|\hat\rho|\mathbf{y}\rangle
\end{equation}
for short and long length-scales respectively, so that the
localization effect saturates.  It is to be pointed out that the
quantity which actually distinguishes the two regimes is of the form
${\bbox{q}}\cdot{\hat {\bbox{x}}}$, where ${\bbox{q}}$ is a typical
value of momentum transfer corresponding to the relevant scattering
dynamics, thus depending on details of microsystem, environment and
their interaction potential, while ${\hat {\bbox{x}}}$ are the
position operators for the microsystem, thus depending on the considered matrix
element. The result obtained by Gallis and
Fleming is
\begin{equation}
   \label{eq:9}
        GF [\hat \rho]  
        =
        \int         d^3 \!{\bbox{q}} d^3 \! {\bbox{q}'}\,
        \frac{g (q)}{2 q^4}\delta (q-q')|f (\bbox{q},\bbox{q}')
        |^2 
        \left (e^{ {i\over\hbar} ({\bbox{q}}-{\bbox{q}}')\cdot{\hat
    {\bbox{x}}}}\hat\rho e^{-{i\over\hbar}({\bbox{q}}-{\bbox{q}}')\cdot{\hat
    {\bbox{x}}}}
          -\hat\rho
          \right ),
\end{equation}
with $g (q)=n (q)v (q)$, where $n(q)$ is the number density of
scattering particles with momentum $q$, $v (q)$ their speed and $f
(\bbox{q},\bbox{q}')$ the scattering amplitude.  Considering the
expression of $GF$ one can check that it actually leads to results
analogous to~(\ref{eq:7}).  Indeed the results~(\ref{eq:9})
and~(\ref{eq:1}) are considered as a standard reference for the study
of decoherence, and they have been recently
exploited~\cite{AlickiPRA,Viale} in trying to quantitatively estimate
decoherence in interference experiments with fullerene molecules.
In~\cite{AlickiPRA} the connection between the models
in~\cite{Joos-Zeh,Gallis90} and the theory of dynamical semigroups is
considered, and the ME
\begin{equation}
   \label{eq:11}
        A [\hat \rho]  
        =
        \int   d^3 \!\bbox{k} \, \tau (\bbox{k})
        \left (e^{ {i\over\hbar} \bbox{k}\cdot{\hat
    {\bbox{x}}}}\hat\rho e^{-{i\over\hbar}\bbox{k}\cdot{\hat
    {\bbox{x}}}}
          -\hat\rho
          \right )
\end{equation}
is proposed, where $\tau (\bbox{k})$ is the density of collisions per
unit time leading to a momentum transfer $\bbox{k}$. The operator
structure and the role of the momentum transfer in its determination
is here put in major evidence. This result can be easily connected
to~(\ref{eq:9}) by observing that setting $|f
({\bbox{q}},{\bbox{q}'})|=|f ({\bbox{q}}-{\bbox{q}'})|\equiv|f
(\bbox{k})|$ one has $\tau ({\bbox{k}})\equiv |f ({\bbox{k}})|^2 \int
\frac{d^3 \!{{\bbox{q}}}}{2 q^4} \, g (q)\delta
(q-|{\bbox{q}}-{\bbox{k}}|)$.  In both~(\ref{eq:9}) and~(\ref{eq:11})
only the position operators ${\hat {\bbox{x}}}$ appear, showing up in
the typical expression $e^{ {i\over\hbar} \bbox{q}\cdot{\hat
    {\bbox{x}}}}$, this unitary operator being strictly related to TI,
as we shall see in Sec.~\ref{sec:covar-prop}.
\par
The absence of the momentum operator in~(\ref{eq:9}) and (\ref{eq:11})
indicates that neither can describe the approach to thermal equilibrium,
and in fact similar to~(\ref{eq:1}) they both predict a steady
growth in energy for the microsystem, neglecting the effect of recoil
in collisions. If only small momentum transfers are of relevance, or
one assumes ${\hat \rho}$ diagonal enough in position representation,
the unitary operators $e^{ {i\over\hbar} \bbox{q}\cdot{\hat
    {\bbox{x}}}}$ can be expanded up to second order, leading from
~(\ref{eq:9}) or~(\ref{eq:11}) to~(\ref{eq:1}), where typical
structures of double commutators with the position operators appear,
corresponding to a Gaussian, diffusive behavior. ME like~(\ref{eq:1})
or~(\ref{eq:2}) can all be obtained starting from the general Lindblad
structure $L[\hat\rho]= \sum_i [ {\hat {V}}_i {\hat \rho} {\hat
  {V}}_i^{\dagger} - \frac 12 \{ {\hat {V}}_i^{\dagger}{\hat {V}}_i,
{\hat \rho} \} ]$ and making the Ansatz: ${\hat V}_i=\alpha_i{\hat
  {{\bf p}}} +\beta_i{\hat {{\bf x}}}$~\cite{SandulescuIJMPE}.
\par
To cope with friction and thermalization to a suitable stationary
state one has to modify~(\ref{eq:9}) or~(\ref{eq:11}) in order to let
the momentum operators of the microsystem ${\hat {{\bf p}}}$ appear,
similar to the modification in going from~(\ref{eq:1})
to~(\ref{eq:2}).  The correction must be such that one has a suitable
thermal stationary state and that energy of the microsystem does not
grow to infinity.  A first significant step in this direction has been
done by Gallis~\cite{Gallis93} with a phenomenological approach which
always takes as starting point the formal Lindblad structure, but
rather than the previous Ansatz assumes the more general expression $
{\hat V} (\mathbf{q})=\alpha (q) e^{ {i\over\hbar} \bbox{q}\cdot{\hat
    {\bbox{x}}}}+\beta (q)e^{ {i\over\hbar} \bbox{q}\cdot{\hat
    {\bbox{x}}}} \bbox{q}\cdot{\hat {\bbox{p}}}$, substituting the sum
over $i$ with an integral over the momentum $\mathbf{q}$, already
putting into evidence the unitary operator $e^{ {i\over\hbar}
  \bbox{q}\cdot{\hat {\bbox{x}}}}$ which played such an important role
in~(\ref{eq:9}) and~(\ref{eq:11}). The result is:
\begin{multline}
   \label{eq:16}
        G [\hat \rho]  
        =
        \int   d^3 \!{\bbox{q}}\, |\alpha (q) |^2
        \left (e^{ {i\over\hbar} {\bbox{q}}\cdot{\hat
    {\bbox{x}}}}\hat\rho e^{-{i\over\hbar}{\bbox{q}}\cdot{\hat
    {\bbox{x}}}}
          -\hat\rho
          \right )   
\\
\nonumber
        +
        \int   d^3 \!{\bbox{q}}\, |\beta (q) |^2
        \left (e^{ {i\over\hbar} {\bbox{q}}\cdot{\hat
    {\bbox{x}}}} {\bbox{q}}\cdot{\hat
    {\bbox{p}}}\hat\rho{\bbox{q}}\cdot{\hat
    {\bbox{p}}} e^{-{i\over\hbar}{\bbox{q}}\cdot{\hat
    {\bbox{x}}}}
          - \frac{1}{2}\{({\bbox{q}}\cdot{\hat
    {\bbox{p}}})^2,\hat\rho\}
          \right )   
\\
\nonumber
        -
        \int   d^3 \!{\bbox{q}} \,
        e^{ {i\over\hbar} {\bbox{q}}\cdot{\hat
    {\bbox{x}}}} \left (
\Re [\alpha^* (q)\beta (q) ] \{{\bbox{q}}\cdot{\hat
    {\bbox{p}}},\hat\rho\} + 
\Im [\alpha^* (q)\beta (q) ][{\bbox{q}}\cdot{\hat
    {\bbox{p}}},\hat\rho]\right ) 
e^{-{i\over\hbar}{\bbox{q}}\cdot{\hat
    {\bbox{x}}}},
\end{multline}
and under certain restrictions on the phenomenological functions
$\alpha(q)$ and $\beta (q)$ does in fact predict relaxation to thermal
equilibrium. Further work in this direction has been done by
Di\'osi~\cite{DiosiEL95} starting from an analogy with the
classical linear Boltzmann equation. He tried to connect similar
structures of ME, in which both position and momentum operator of the
microsystem appear, to an underlying dynamics in terms of collisions obtaining
the result
\begin{equation}
   \label{eq:17}
        D [\hat \rho]  
        =
        \frac{n m^3}{\mu^5}
        \int         d^3 \!{{\bbox{q}}}d^3 \!{{\bbox{q}}'}\,
        \delta (E (q) -E(q'))|f ({\bbox{q}},{\bbox{q}}')|^2 
        \left( 
        {\hat V}  \hat\rho{\hat V}^{\dagger} -
        \frac{1}{2}\{ {\hat V}^{\dagger}{\hat V},\hat\rho\}
        \right)
\end{equation}
with ${\hat V} = \sqrt{\sigma (\mathbf{q}+\frac{m}{M} ({\hat
    {\bbox{p}}}+\mathbf{q}))} e^{ {i\over\hbar}
  ({{\bbox{q}}}-{{\bbox{q}}}')\cdot{\hat {\bbox{x}}}}$, $M$ mass of
the microsystem, $m$ the mass of the gas particles, $\mu$ the reduced
mass, $n$ the gas density, $E (q)=\frac{q^2}{2M}$ and $\sigma$ the
momentum distribution of the gas particles. If $\sigma$ is given by a
Boltzmann distribution an operator of the form $e^{-\beta { {\hat
      {\bbox{p}}}^2 \over 2M } }$ is a stationary solution
of~(\ref{eq:17}). A general result for a ME describing the motion of a
particle interacting through collisions with some surrounding
environment has been recently obtained starting from a scattering
theory derivation~\cite{art5,art3-art4-art7-art6}. The result relies
on the appearance of a two-point correlation function known as dynamic
structure factor, operator valued due to its dependence on the
momentum operators of the microsystem. The dynamic structure factor
obeys the detailed balance condition and therefore grants the
existence of the expected stationary solution on very general grounds.
The ME is
\begin{displaymath}
        V [\hat \rho]  
        \!=\!
        (2\pi)^4 \hbar^2 n \!\!
        \int d^3\!
        {{\bf q}}
        \,  
        {
        | \tilde{t} (q) |^2
        }
      \Biggl[
        e^{{i\over\hbar}{{\bf q}}\cdot{\hat {{\bf x}}}}
        \sqrt{
        S({{\bf q}},{\hat {{\bf p}}})
        }
        {\hat \rho}
        \sqrt{
        S({{\bf q}},{\hat {{\bf p}}})
        }
        e^{-{i\over\hbar}{{\bf q}}\cdot{\hat {{\bf x}}}}
        -
        \frac 12
        \left \{
        S({{\bf q}},{\hat {{\bf p}}}),
        {\hat \rho}
        \right \}
        \Biggr],
\end{displaymath}
with $\tilde{t} (q)$ Fourier transform of the T-matrix
describing the microphysical collisions and $S({\bf{q}},
{\bf{p}})$ the positive two-point correlation function
\begin{displaymath}
  S({\bf{q}},E) = 
        {  
        1  
        \over  
         2\pi\hbar
        }  
        \int  dt 
        {\int d^3 \! {\bf{x}} \,}        
        e^{
        {
        i
        \over
         \hbar
        }
        [E ({\bf{q}},{\bf{p}}) t -
        {\bf{q}}\cdot{\bf{x}}]
        } 
      \frac{1}{N}
        {\int d^3 \! {\bf{y}} \,}
        \left \langle  
         N({\bf{y}})  
         N({\bf{y}}+{\bf{x}},t)  
        \right \rangle    
\end{displaymath}
with $S({\bf{q}},{\bf{p}})\equiv S({\bf{q}},E)$, $E
({\bf{q}},{\bf{p}})= { ({\bf{p}}+{\bf{q}})^2 \over 2M } - { {\bf{p}}^2
  \over 2M }$,
${\bf{q}}$ and $E$ being momentum and energy transfer, while
$N({\bf{y}})$ is the particle density operator in the environment.
\section{Covariance properties}
\label{sec:covar-prop}
The validity of a ME for the description of the reduced
dynamics of a microsystem interacting with some environment ultimately
rests on how realistic the environment and its coupling to the quantum
system of interest have been described and how severe the
approximations allowing for the derivation of the ME for
the reduced system actually are. It is nevertheless of interest,
and of guidance in determining equations giving the time evolution of
the statistical operator, to check whether some general features, which
should be common to any dynamical evolution, are actually
present. Among these features one has preservation of trace and
positivity of the statistical operator; complete positivity which
emerges as a typical feature of quantum mechanics related to the non
commutativity of the algebra of observables~\cite{HolevoNEW};
preservation of typical symmetries of the environment such as
homogeneity~\cite{Tannor97} and in general invariance under the action
of a group expressing some symmetry of the whole physical system;
existence and uniqueness of a suitable stationary state with a
canonical structure; correct description of the time evolution of the
observables relevant to the dynamics, such as energy.
\par
The most widespread approach is to start from or compare with the
Lindblad structure of a ME, both in presence of bounded and unbounded
operators, so that complete positivity and therefore in particular
positivity is granted, and the same goes for preservation of the
trace. In the case in which the physical system is characterized by
some non trivial symmetry group however, one can rely on more recent
and refined results than the one by Lindblad. The possible structures
of generators of quantum dynamical semigroups covariant under the
action of a symmetry group have been characterized in particular in
the case of the two Abelian Lie groups $\mathrm{R}$ and
$\mathrm{U(1)}$~\cite{HolevoRAN-HolevoJMP}, also taking care of
defining a suitable domain in the case in which the relevant operators
are unbounded.
\par
Since we are focusing on structures of ME
describing the loss of coherence of a microsystem interacting through
collisions with a homogeneous environment, we will consider in some
detail only the structure of the generator of a TI
quantum dynamical semigroup. This result has been settled by Holevo
and gives a non-commutative quantum generalization of the
L\'evy-Khinchin formula. In the following we will try to briefly
summarize Holevo's results.
\par
Let us first consider the case of a norm-continuous conservative
quantum dynamical semigroup $\{\Phi_t; t\geq 0\}$ acting on the
algebra of bounded operators in $L^2 ({\bf R}^{3})$, whose generator
$\mathcal{L}$ is a completely dissipative map satisfying
\begin{equation}
   \label{eq:21}
   \frac{d}{dt}\Phi_t[{\hat X}]={\cal L}[\Phi_t[{\hat X}]]\quad  X\in{\cal B}
  (L^2 ({\bf R}^{3})) \quad t\geq 0
\end{equation}
with $\Phi_0[{\hat X}]={\hat X}$ and $\Phi_t[{\hat I}]={\hat I}$ due to
conservativity, i.e. trace preservation. If $\Phi_t$ is
norm-continuous the generator $\mathcal{L}$ admits a \textit{standard
  representation}
\begin{equation}
   \label{eq:22}
   {\cal L}[{\hat X}]=
        {i \over \hbar}
        [{\hat H}
        ,
        {\hat \rho}
        ]
        +
 \Psi[{\hat X}]-\frac{1}{2}\{\Psi[{\hat I}],{\hat X}\}
\quad  {\hat X}\in{\cal B}
  (L^2 ({\bf R}^{3})) 
\end{equation}
with $\Psi$ a normal completely positive map and ${\hat H}$
self-adjoint. The semigroup is said to be covariant under the action
of a unitary representation $\hat{U} ({{\bbox{q}}})=e^{{i\over\hbar}{{\bbox{q}}}\cdot{\hat
    {\bbox{x}}}}$, ${{\bbox{q}}} \in {\bf R}^3$ of the group of
translations, i.e. translation-covariant, provided
\begin{equation}
   \label{eq:23}
   \Phi_t[\hat{U}^{\dagger} ({{\bbox{q}}}){\hat X}\hat{U}
   ({{\bbox{q}}})]=\hat{U}^{\dagger} ({{\bbox{q}}})\Phi_t[{\hat X}]\hat{U}
   ({{\bbox{q}}}) \quad
    {\hat X}\in{\cal B}
  (L^2 ({\bf R}^{3})) \quad {{\bbox{q}}} \in {\bf R}^3 \quad t\geq 0
\end{equation}
holds. If $\mathcal{L}$ is covariant in the sense of~(\ref{eq:23}) in
the decomposition~(\ref{eq:22}) the map $\Psi$ can always be
chosen covariant and ${\hat H}$  commuting with the unitary
representation $\hat{U} ({{\bbox{q}}})$. Under the
restriction~(\ref{eq:23}) the general structure of the bounded
generator of the semigroup in~(\ref{eq:21}) is in fact given by
\begin{multline}
   \label{eq:24}
\!\!\!\!\!  {\cal L}[{\hat X}]={i \over \hbar}
        \left[
        H ({\hat {{\bf p}}})
        ,{\hat X}
        \right]
\\
\! +\!
\int \!\sum_{j=1}^{\infty}
\left[
L^{\dagger}_j({{\bbox{q}}},{\hat {\bbox{p}}})
\hat{U}^{\dagger} ({{\bbox{q}}}){\hat X}\hat{U} ({{\bbox{q}}})
L_j({{\bbox{q}}},{\hat {\bbox{p}}})
\!-\!
        \frac 12
        \left \{
        L^{\dagger}_j({{\bbox{q}}},{\hat
          {\bbox{p}}})L_j({{\bbox{q}}},{\hat {\bbox{p}}}),{\hat X} 
        \right \}
\right]
d\mu ({{\bbox{q}}})  
\end{multline}
where $H (\cdot)=H^{*} (\cdot)$, $L_j({{\bbox{q}}},\cdot)$ are bounded
functions, $\mu ({{\bbox{q}}})$ is a positive $\sigma$-finite measure on
${\bf R}^3$ and  $\int \sum_{j=1}^{\infty}
  |L_j({{\bbox{q}}},\cdot)|^2d\mu ({{\bbox{q}}})<+\infty$.
\par
In the case in which the family of maps $\{\Phi_t; t\geq 0\}$
acting on ${\cal B} (L^2 ({\bf R}^{3})) $ are generally unbounded it
is convenient to consider the equation
\begin{displaymath}
   \frac{d}{dt}\langle\phi|\Phi_t[{\hat X}]\psi\rangle={\cal L}
(\phi;\Phi_t[{\hat X}];\psi)\quad {\hat X}\in{\cal B}
  (L^2 ({\bf R}^{3})) \quad t\geq 0
\end{displaymath}
with $\Phi_0[{\hat X}]={\hat X}$, $\Phi_t[{\hat I}]={\hat I}$ and
$\phi,\psi\in {\cal D}\subset L^2 ({\bf R}^{3})$, where ${\cal D}$ is
some dense domain. The expression ${\cal L} (\phi;{\hat X};\psi)$ is
the so-called \textit{form-generator}, i. e. a function of
$\phi,\psi\in {\cal D}\subset L^2 ({\bf R}^{3})$ and ${\hat X}\in{\cal
  B} (L^2 ({\bf R}^{3}))$ characterized by the following basic
properties: 1) ${\cal L} (\phi;{\hat X};\psi)$ is linear in ${\hat X}$
and $\psi$, anti-linear in $\phi$ and such that ${\cal L}^*
(\phi;{\hat X};\psi)={\cal L} (\psi;{\hat X}^{\dagger};\phi)$; 2) for
all finite subsets $\{\psi_j\}\in {\cal D}\subset L^2 ({\bf R}^{3})$
and $\{{\hat X}_j\}\in{\cal B} (L^2 ({\bf R}^{3}))$ such that $\sum_j
{\hat X}_j\psi_j=0$ one has $\sum_{jk}{\cal L} (\psi_j;{\hat
  X}^{\dagger}_j{\hat X}_k;\psi_k)\geq 0$ (conditional complete
positivity); 3) ${\cal L} (\phi;{\hat I};\psi)=0$ $\forall
\phi,\psi\in {\cal D}\subset L^2 ({\bf R}^{3})$ (conservativity);
together with suitable continuity properties. The form-generator also
admits a \textit{standard representation}
\begin{displaymath}
  {\cal L}(\phi;{\hat X};\psi)\!=\!
  \sum_j \langle {\hat L}_j\phi|{\hat X}{\hat L}_j\psi\rangle 
-\langle {\hat K} \phi|{\hat X}\psi\rangle-\langle\phi|{\hat X}{\hat K}\psi\rangle
   \quad\! \phi,\psi\in
{\cal D},
\ {\hat X}\in{\cal B} (L^2 ({\bf
  R}^{3}))
\end{displaymath}
with ${\hat K}$ and ${\hat L}_j$ densely defined operators. The covariance
condition analogous to~(\ref{eq:23}) is now expressed by ${\cal
  L}(\phi;\hat{U}^{\dagger} ({{\bbox{q}}}) {\hat
  X}\hat{U}({{\bbox{q}}});\psi)= {\cal L}(\hat{U}({{\bbox{q}}})\phi;
{\hat X};\hat{U}^{\dagger} ({{\bbox{q}}})\psi)$, together with the
invariance of the domain ${\cal D}$ under the unitary representation.
It is however no longer possible to put into evidence an Hamiltonian
contribution commuting with the unitary representation and a
completely positive map. Taking as domain the space of twice
continuously differentiable functions with compact support in the
momentum representation of the CCR, i.e. ${\cal D}=C^{2}_{0}({\bf
  R}^{3})$, and asking for suitable continuity properties the general
structure of the TI form-generator is given by $ {\cal L}={\cal
  L}_{G}+{\cal L}_{P}$, where ${\cal L}_{G}$ is the Gaussian,
continuous component corresponding to the formal operator expression
\begin{gather}
   \label{eq:29}
{\cal L}_{G}[{\hat X}]={i \over \hbar}
        \left[{\hat {\bbox{y}}}_0+
        H ({\hat {{\bf p}}})
        ,{\hat X}
        \right]
+
\sum_{k=1}^{3}
\left({\hat {\bbox{y}}}_k+L_k ({\hat {\bbox{p}}}) \right)^{\dagger}
{\hat X}
\left({\hat {\bbox{y}}}_k+L_k ({\hat {\bbox{p}}}) \right)-{\hat
  K}^{\dagger}{\hat X}-{\hat X}{\hat K}
\nonumber\\
K=\frac{1}{2}\sum_{k=1}^{r}\left({\hat {\bbox{y}}}_k^2+2{\hat
      {\bbox{y}}}_kL_k ({\hat {\bbox{p}}}) +L^{\dagger}_k ({\hat
      {\bbox{p}}})L_k ({\hat {\bbox{p}}}) \right)
\end{gather}
with ${\hat {\bbox{y}}}_k=\sum_{i=1}^{3}a_{ki}{\hat {\bbox{x}}}_i$, 
$k=0,\ldots,3$, $a_{ki}\in {\bf R}$, 
$H (\cdot)=H^{*} (\cdot)\in
L^2_{{\rm \scriptscriptstyle loc}} ({\bf R}^{3})$ and $\left|L_k(\cdot)\right|^2\in
L^2_{{\rm \scriptscriptstyle loc}} ({\bf R}^{3})$, while ${\cal
  L}_{P}$ is the Poisson, jump component
\begin{multline}
   \label{eq:30}
\!\!\!\!\!{\cal L}_{P}[{\hat X}]\!=\!\!
\int \!\sum_{j=1}^{\infty}
\!\left[
L^{\dagger}_j({{\bbox{q}}},{\hat {\bbox{p}}})
\hat{U}^{\dagger} ({{\bbox{q}}}){\hat X}\hat{U} ({{\bbox{q}}})
L_j({{\bbox{q}}},{\hat {\bbox{p}}})
    \!   -\!
        \frac 12
        \left \{
        L^{\dagger}_j({{\bbox{q}}},{\hat
          {\bbox{p}}})L_j({{\bbox{q}}},{\hat {\bbox{p}}}),{\hat X} 
        \right \}\!
\right]\!
d\mu ({{\bbox{q}}})
\\
+
\int \sum_{j=1}^{\infty}
\left[
\omega_j ({{\bbox{q}}})
L^{\dagger}_j({{\bbox{q}}},{\hat {\bbox{p}}})
(\hat{U}^{\dagger} ({{\bbox{q}}}){\hat X}\hat{U} ({{\bbox{q}}})-{\hat X})
\right.
        \\
\left.
        \hphantom{pippopippopippopippo}
{}+
(\hat{U}^{\dagger} ({{\bbox{q}}}){\hat X}\hat{U} ({{\bbox{q}}})-{\hat X})
L_j({{\bbox{q}}},{\hat {\bbox{p}}})\omega_j^{*} ({{\bbox{q}}})
\right]
d\mu ({{\bbox{q}}})
\\
+
\int \sum_{j=1}^{\infty}
\left[
\vphantom{-i
\sum_{k=1}^{r}\frac{[{\hat X},{\hat {\bbox{x}}}_k]}{1+|{{\bbox{q}}}|^2}{{\bbox{q}}}_k}
\hat{U}^{\dagger} ({{\bbox{q}}}){\hat X}\hat{U} ({{\bbox{q}}})
-{\hat X}
-i \frac{[{\hat X},{\hat {\bbox{x}}}\cdot{\bbox{q}}]}{1+|{\bbox{q}}|^2}
\right]
|\omega_j ({{\bbox{q}}})|^2
d\mu ({{\bbox{q}}})
\end{multline}
with $\mu ({{\bbox{q}}})$ a positive $\sigma$-finite measure on ${\bf
  R}^3$, $\omega_j ({{\bbox{q}}})$ complex measurable functions and
the further conditions $\int\, {|{{\bbox{q}}}|^2}/
({1+|{{\bbox{q}}}|^2}) \sum_{j=1}^{\infty}|\omega_j
({{\bbox{q}}})|^2d\mu ({{\bbox{q}}})<+\infty$ and $\int\,
\sum_{j=1}^{\infty}|L_j({{\bbox{q}}},\cdot)|^2 d\mu ({{\bbox{q}}})\in
L^2_{{\rm \scriptscriptstyle loc}} ({\bf R}^{3})$,
equations~(\ref{eq:29}) and~(\ref{eq:30}) giving a non-commutative
quantum generalization of the L\'evy-Khinchin formula. Despite
appearance the result can still be cast in Lindblad form.
\par
All the ME for the statistical operator considered in
the previous section can be formally compared to the pre-adjoint of the
maps given in~(\ref{eq:24}) or~(\ref{eq:29}) and~(\ref{eq:30}),
with suitable choices of parameters and
functions~\cite{art5,art8}. One thus sees that not all the
ME proposed in Sec.~\ref{sec:models-decoh-induc} are
proper generators of quantum dynamical semigroups, as already
mentioned in connection with the property of complete positivity. In
particular one can now clearly distinguish between Gaussian and
Poisson components. Eq.~(\ref{eq:29}) and~(\ref{eq:30}) also
give some hints about the possible structures of the ME
for a microsystem interacting with a homogeneous environment which
might be derived in future research work, starting form detailed
physical models. Of course the case of a microsystem interacting
through collisions with a TI bath is just one of
the possible physical models interesting for the study of decoherence, 
another most important example is the interaction of a charged
particle with the electromagnetic field and the related phenomenon of
decoherence due to Bremsstrahlung~\cite{PetruccioneBREMS}.
\par
\textit{Note added}. After completion of the first version of the
manuscript, further work deserving attention has been done on the
subject~\cite{HalliwellQBM-ZeilingerQBM-th}, though from a different
standpoint.

\end{document}